# Interfacial Multiferroics of TiO$_2$/PbTiO$_3$ Heterostructure Driven by Ferroelectric Polarization Discontinuity

Fang Wang[1], Zhaohui Ren[1], He Tian[1,2], Shengyuan A. Yang[3], Yanwu Xie[4], Yunhao Lu[1*], Jianzhong Jiang[1], Gaorong Han[1] and Kesong Yang[5]

[1] State Key Laboratory of Silicon Materials, School of Materials Science and Engineering, Zhejiang University, Hangzhou, 310027, China

[2] Center of Electron Microscope, School of Material Science & Engineering, Zhejiang University, Hangzhou, 310027, China

[3] Research Laboratory for Quantum Materials, Singapore University of Technology and Design, Singapore 487372, Singapore

[4] Department of Physics, Zhejiang University, Hangzhou, 310027, China

[5] Department of NanoEngineering, University of California, San Diego, 9500 Gilman Drive, Mail Code 0448, La Jolla, California 92093-0448, USA

**Abstract:**

Novel phenomena appear when two different oxide materials are combined together to form an interface. For example, at the interface of $LaAlO_3/SrTiO_3$, two dimensional conductive states form to avoid the polar discontinuity and magnetic properties are found at such interface. In this work, we propose a new type of interface between two nonmagnetic and nonpolar oxides that could host a conductive state with magnetic properties, where it is the ferroelectric polarization discontinuity instead of the polar discontinuity that leads to the charge transfer, forming the interfacial conductive or magnetic states. As a concrete example, we investigate by first-principles calculations the heterostructures made of ferroelectric perovskite oxide $PbTiO_3$ and non-ferroelectric polarized oxides $TiO_2$. We show that charge is transferred to the interfacial layer forming an interfacial conductive state with ferromagnetic ordering that may persist up to room temperature. Especially, the strong coupling between bulk ferroelectric polarization and interface ferromagnetism represents a new type of magnetoelectric effect, which provides an ideal platform for exploring the intriguing interfacial multiferroics. The findings here are important not only for fundamental science but also for promising applications in nanoscale electronics and spintronics.

Perovskite oxides are a class of materials hosting many excellent properties such as ferroelectricity, piezoelectricity, and ferromagnetism. More interestingly, novel phenomena appear when two different oxide materials are combined together to form an interface. As a canonical example, the polar/nonpolar $LaAlO_3/SrTiO_3$ heterostructure has been receiving immense interest over the past decade.[1] In this heterostructure system, although both constituent materials are wide-gap band insulators, the interface becomes conducting with a confined two-dimensional (2D) electronic state.[1,2] The formation of interfacial conductive state can be attributed to the polar discontinuity at the n-type $(LaO)^+/(TiO_2)^0$ interface. Along the (001) growth direction, $LaAlO_3$ consists of alternating stacked polar $(LaO)^+$ and $(AlO_2)^-$ layers, whereas $SrTiO_3$ is composed of nonpolar $(SrO)^0$ and $(TiO_2)^0$ layers. At the interface, to avoid the polar discontinuity, one half electron is transferred from $(LaO)^+$ to the interfacial $TiO_2$ layer, forming the interfacial conducting states. Besides the conductive state, the oxide interface also provides a versatile platform to generate and manipulate 2D magnetic ordering. For example, the ferromagnetism was introduced by delta doping $EuTiO_3$ at the interface and the spin-polarization is fully electric-field-tunable.[3] And the switching of magnetic domains was also found at $LaAlO_3/SrTiO_3$ interface.[4] However, the intrinsic ferromagnetism at $LaAlO_3/SrTiO_3$ interface is very weak[5,6] and the true long-range magnetic ordering at this interface is not well-established and still under debate, which may be related to treatment-dependent impurity.[7-9] The interfacial magnetism is determined by the delicate coupling between spin, orbital, charge and lattice degrees of freedom and developing a new effective strategy other than polar discontinuity to engineer these intricate couplings in oxide heterostructures is needed to realize stable interfacial magnetism with robust long-range magnetic ordering.

With the rapid development in the atomic-scale synthesis and characterization techniques, various high-quality oxide interfaces can be fabricated, which motivates the exploration of new types of oxide interfaces that could exhibit magnetism. So far, most attention has been paid on the heterostructures involving polar materials.[10] In principle, an interfacial conductivity could also be generated at the interface between two *nonpolar* oxides with and without ferroelectric or strain-induced polarization[11, 12], where it is the ferroelectric polarization discontinuity instead of the polar discontinuity that leads to the charge transfer, forming the interfacial conductive states.[13] Similarly, the interfacial magnetism can be driven by the same mechanism. This strategy gains support by the fact that many perovskite oxides are excellent ferroelectric (piezoelectric) materials, and it has the great advantage that the interfacial electric and/or magnetic properties may be readily manipulated by controlling the polarization through an electric field. A few candidate heterostructure systems consisting of ferroelectric and paraelectric materials have been proposed recently, such as $BaTiO_3/SrTiO_3$[14] and $PbTiO_3/SrTiO_3$[15] heterostructures. However, the interfacial magnetism driven by ferroelectric polarization discontinuity in these systems is still elusive and spin-polarized carriers are found only at those interfaces with magnetic oxides where the interfacial magnetism derives from the bulk magnetic ordering.[16-18]

In this work, we propose a new type of interface between two nonmagnetic and nonpolar oxides that could host a conductive state with ferromagnetism. As a concrete example, we investigate the heterostructure made of perovskite oxide $PbTiO_3$ and anatase $TiO_2$ using first-principles calculations. $PbTiO_3$ is a typical ferroelectric material with large polarization strength and has electron-balanced alternating layers without polar planes. Although anatase $TiO_2$ is not of the perovskite structure, it also has the edge-shared oxygen octahedra and its lattice constant of the (001) facet matches well with that of $PbTiO_3$ in (001) surface. We show that a high-quality interface can be formed between the two materials, and due to the ferroelectric polarization discontinuity at the interface, charge is transferred to the interfacial ($TiO_2$) layer forming an interfacial conductive state with strong ferromagnetic ordering that may persist up to room temperature. More importantly, we show that the interfacial magnetism is sensitive to the ferroelectric polarization strength of $PbTiO_3$, hence can be readily tuned through an applied electric field. Especially, the strong coupling between bulk ferroelectric polarization and interface ferromagnetism represents a new type of magnetoelectric effect, which provides an ideal platform for exploring the intriguing interfacial multiferroics.[19] The effects of strain on the interfacial properties are also discussed. To our best knowledge, this is the first time that an interfacial magnetism is found at a nonmagnetic nonpolar/nonpolar or ferroelectric/paraelectric oxide interface, which opens a new route for achieving controllable 2D spin-polarized carriers. The findings here are important not only for fundamental science but also for promising applications in nanoscale electronics and spintronics.[20]

At room temperature (below 763 K), PbTiO$_3$ crystallizes in a tetragonal phase with space group No. 99 (P4*mm*). The calculated PbTiO$_3$ lattice constants $a$ = 3.89 Å and $c/a$ = 1.04 are well comparable with the experimental values[21] of $a$ = 3.90 Å and $c/a$ = 1.06, and the previously reported theoretical values[22] of $a$ = 3.87 Å and $c/a$ = 1.042. Its structure consists of alternating charge-neutral (TiO$_2$)$^0$ and (PbO)$^0$ planes along the [001] direction, which is analogous to the nonpolar material SrTiO$_3$. However, different from SrTiO$_3$, a spontaneous ferroelectric polarization occurs in PbTiO$_3$ due to the off-center displacement of the cation. The optimized displacement between O and Pb/Ti along the [001] direction is about 0.32 Å, resulting in a large ferroelectric polarization. The calculated polarization, $P_{\text{PTO}}$=0.72C/m$^2$ based on Born effective charges method, agrees well with experimental value (~0.75C/m$^2$)[23] and the result using the Berry phase method (~0.77C/m$^2$).[15] For TiO$_2$ in the anatase phase, our calculated lattice constants $a$ and $c$ are 3.78 Å and 9.49 Å respectively, in good agreement with the previously reported experimental[24] and theoretical values.[25] In constructing the heterostructure model, the in-plane lattice constant of the superlattice is fixed to that of PbTiO$_3$. This results in a 3% tensile strain on the lateral size of TiO$_2$, which has no apparent effect on the electronic structure of TiO$_2$ except for a little narrowing of its bandgap. As TiO$_2$ can be viewed as distorted (TiO$_2$)$^0$ planes stacking along the [001] direction without either polar or ferroelectric polarization, the polarization discontinuity emerges at the TiO$_2$/PbTiO$_3$(001) interface. To be specific, in our modeling, we firstly take the ferroelectric polarization of PbTiO$_3$ to be oriented from PbTiO$_3$ to TiO$_2$, as shown in Figure 1(a), corresponding to an n-type interface.

The calculated layer-resolved total density of states (DOS) (summing both spin majority and minority channels) and partial DOS (PDOS) projected on O-$2p$ and Ti-$3d$ orbitals around the interface are shown in Figure 1(b). As expected, a conductive state does emerge at the interface between the two nonpolar materials. One can clearly observe that the ferroelectric polarization of PbTiO$_3$ drives a charge transfer to the interfacial TiO$_2$ layer (IF-TiO$_2$), forming the interfacial conductive states. It is noted that in practical case, this charge transfer or ferroelectric polarization is stabilized by capped metal gate, surface reconstruction or adsorbed molecules. The interfacial conductive states near the Fermi level are highly concentrated at the IF-TiO$_2$ layer which is shared by both PbTiO$_3$ and TiO$_2$, with only slight spread to the neighboring two or three layers on each side. All the other layers away from the interface exhibit insulating behavior, having similar electronic structures with that of the bulk TiO$_2$ or PbTiO$_3$. All these characteristics show that a 2D conductive state forms at the oxide interface. As the Ti-$3d$ orbitals are the lowest unoccupied states in bulk TiO$_2$ and the (TiO$_2$) layer of PbTiO$_3$, the conductive states at IF-TiO$_2$ are mostly contributed by the Ti-$3d$ orbitals. These partially occupied and unpaired $d$ electrons form a spin-polarized conductive state at the interface (see Figure 2(a)). In the octahedral crystal field, the three degenerated $t_{2g}$ orbitals have a lower energy. The interfacial TiO$_6$ octahedral distortion further splits the three $t_{2g}$ orbitals: the $d_{xy}$ orbital has a lower energy and is much more dispersive in plane than the $d_{xz}$ and $d_{yz}$ orbitals (Figure 2(c)). As a result, the $d_{xy}$ orbital is firstly occupied and almost solely responsible for the interfacial spin-polarized metallic states, a scenario similar to the case of the LaAlO$_3$/SrTiO$_3$ interface.[26] This picture is further supported by the $d_{xy}$-character-like orbital shape in the spin density plot as shown in Figure 2(b). The $d_{yz}$ and $d_{xz}$ orbitals stay at higher energies in the conduction band and are less occupied (Figure 2(c)).

As discussed above, the ferroelectric-polarization-discontinuity drives the charge transfer from the PbTiO$_3$ to the TiO$_2$ layer, and these transferred electrons partially occupied Ti 3$d$ orbitals, leading to the formation of the conductive state at the interface. These unpaired Ti 3$d$ electrons at the IF-TiO$_2$ layer also induces magnetic moments. For a practical application purpose, a stable ferromagnetic ordering between these magnetic moments is desired. One may suspect that the magnetic coupling between two adjacent Ti$^{3+}$ ($d^1$) ions is antiferromagnetic,[27] as in the case of the bulk LaTiO$_3$.[28] However, in the TiO$_2$/PbTiO$_3$ heterostructure, the transferred electrons are far less than one e per Ti ion, and thus the magnetic coupling between the Ti ions at the TiO$_2$/PbTiO$_3$ interface may give rise to a totally different picture resulting in a ferromagnetic ordering.

The stability of a ferromagnetic ordering can be further evaluated by the Curie temperature $T_C$. In a Heisenberg model, $T_C$ is determined by the exchange coupling between the atomic moments on different sites. Although this model is not applicable for itinerant magnetism with partially occupied orbitals, the stability of ferromagnetic (FM) ordering at interface can be inferred from the total energy differences between different magnetic configurations. Figure 3(a) shows the energy differences between FM and antiferromagnetic (AFM) configurations as a function of ferroelectric polarization. Here, the result is shown for the checkerboard AFM ordering, which is more stable than striped AFM ordering in our calculations. We can see that there is no energy difference between FM and AFM ordering when the polarization strength is small and the interface is paramagnetic without persistent spin-polarization. When the ferroelectric polarization increases above ~0.3 C/m$^2$, the conductive state becomes spin-polarized and the FM ordering is energetically more favorable than the AFM ordering. The FM ordering becomes more and more stable than the AFM ordering with further increase in the ferroelectric polarization. The energy difference is as large as 70 meV at the normal ferroelectric polarization of PbTiO$_3$ (0.72 C/m$^2$), which is much larger than that of the most studied LaAlO$_3$/SrTiO$_3$ interface (~1.3 meV[29] and is 8.0 meV from our calculation). This suggests that the FM coupling of TiO$_2$/PbTiO$_3$ interface is stronger than that of the LaAlO$_3$/SrTiO$_3$ interface. The FM ordering of the LaAlO$_3$/SrTiO$_3$ interface has been observed up to room temperature.[30] Since if this magnetic ordering is really intrinsic and exists at the interface, we expect that the FM ordering of TiO$_2$/PbTiO$_3$ can also persist at room temperatures.

The origin of this FM ordering can be understood as a result of the competition between the kinetic energy and the exchange energy of the itinerant carriers in the band model. Spontaneous spin-polarization occurs when the relative gain in exchange interaction is larger than the loss in kinetic energy, which is conventionally formulated as the Stoner criterion[31] D($E_f$)×$J$>1, where D($E_f$) is nonmagnetic DOS at Fermi level and $J$ is the strength of the exchange interaction of the corresponding orbital. The Stoner parameter $J$ can be estimated from the spin-splitting in the FM state,[32] which describes correctly the occurrence of itinerant ferromagnetism in bulk metals.[33] Figure 3(b) shows the nonmagnetic DOS of IF-TiO$_2$ at different ferroelectric polarization strengths of PbTiO$_3$. When the ferroelectric polarization of PbTiO$_3$ is weak, only a small amount of charge transfers to the IF-TiO$_2$ layer for compensating the ferroelectric polarization discontinuity and the occupation of lowest interface Ti 3d states is small, which are mainly of the Ti $d_{xy}$ orbitals. As the Ti $d_{xy}$ orbtial is more dispersive in energy due to octahedral distortion, D($E_f$) is small (~0.5 states/Ti). Combined with an estimated $J$ ~1.7 (Figure.S1), their product is lower than the critical value required by the Stoner criterion hence the interface is nonmagnetic. With increasing polarization strength, the less dispersive $d_{yz}$ and $d_{xz}$ orbitals become occupied and contribute to a sharp increase of the D($E_f$) (see Figure 3(b)). With the normal polarization of PbTiO$_3$ (0.72 C/m$^2$), the Stoner criterion is amply satisfied (middle panel, Figure 3(c)) hence the spontaneous spin polarization occurs for the interfacial states. Although the $d_{yz}$ and $d_{xz}$ orbitals play an important role in the magnetic instability, it is the lower-lying $d_{xy}$ that is mostly occupied after spin-splitting (Figure 2(c)). The Stoner parameter decreases at even larger ferroelectric polarization due to the decreasing of $J$ (Figure.S1). However, it is still larger than the critical value because of the large D($E_f$), resulting in a stable FM state.

From the above discussion, we see that the stability of the ferromagnetism, represented by the Curie temperature, can be controlled by the polarization strength of bulk PbTiO$_3$, representing a new type of magnetoelectric effect. Herein, we further show that, the magnitude of the magnetic moment and the charge carrier density of the conductive state at the interface can also be efficiently tuned by the ferroelectric polarization. In Figure 4(a), we plot the magnetic moment per interface unit cell (uc) as a function of the ferroelectric polarization of PbTiO$_3$. When the polarization is small, the heterostructure is a paramagnetic system with zero magnetic moment. The interfacial magnetic moment of such heterostructure becomes nonzero at the critical polarization as discussed before and varies sensitively with the polarization strength of PbTiO$_3$. The interfacial carrier density is also calculated by integrating the PDOS of the occupied conduction bands at the interface, as shown in Figure 4(b). When the polarization is small, the charge carrier density at the interface is almost zero. This carrier density has a sharp increase above ~0.3 C/m$^2$. This sensitive dependence is partly due to the rigid structure of TiO$_2$ which hardly reconstructs to balance the ferroelectric polarization discontinuity induced by PbTiO$_3$. Thus, strong electronic reconstruction happens in response to the increasing polarization discontinuity and both carrier density and magnetic moment at the interface are highly dependent on ferroelectric polarization, which is also crucial for the occurrence of spin polarization. When the ferroelectric polarization increases, both charge carrier density and magnetic moment at the interface increase and reach ~1.5×10$^{14}$/cm$^2$ and 0.25 μ$_B$/uc respectively at the normal ferroelectric polarization value of PbTiO$_3$. They can be increased further by strain or by external electric field. For example, the polarization of PbTiO$_3$ increases up to ~0.85 C/m$^2$ under 3% compressive strain (Figure.S2) which generates an interfacial carrier density and magnetic moment with ~2.1×10$^{14}$/cm$^2$ and ~0.4 μ$_B$/uc respectively. Therefore, the interfacial carrier density and magnetism are readily tunable by ferroelectric polarization of PbTiO$_3$ and can be tuned in a wide range. The ferroelectric polarization of PbTiO$_3$ is sensitive to the external strain and electric field. For example, the polarization strength varies as large as 20% under in-plane 3% biaxial strain (Figure.S2). Thus, the proposed TiO$_2$/PbTiO$_3$ interface is a promising candidate for realizing interfacial multiferroic effects and 2D spintronics.

When the polarization of PbTiO$_3$ is reversed and switched to be oriented from $TiO_2$ to $PbTiO_3$, as shown in Figure 5(a), the ferroelectric polarization of $PbTiO_3$ drives a charge transfer from the interfacial region. As shown in Figure.5(b), the top of valence bands shifts above Fermi level with some empty states at the IF-$TiO_2$ layer and two neighboring layers ($TiO_2$ and PbO layer). These empty states vanish quickly and insulating bulk states recover away from the interface. All these characteristics show that a 2D p-type conductive state forms at the interface. This is different from the polar/nonpolar $LaAlO_3$/$SrTiO_3$ heterostructure where only electronic (n-type) conductivity is observed at the interface. As the O-*2p* orbitals are the highest occupied states in bulk $TiO_2$ and $PbTiO_3$, the conductive states at interface region are mostly contributed by the O-*2p* orbitals. These partially occupied *2p* states form a spin-polarized conductive state at the interface (see Figure 6(a)). The three *p*-orbitals ($p_x$, $p_y$ and $p_z$) are almost degenerated without large splitting in energy. As a result, all of them are responsible for the interfacial spin-polarized metallic states and orbital shape in the spin density plot as shown in Figure 6(b) also supports this picture.

The origin of this *p*-orbitals spontaneous spin-polarization still can be understood by Stoner Criterion, similar to the hole-induced ferromagnetism in bulk oxides.[34] It has been reported that O-*2p* orbital has very large exchange interaction *J* due to its localized nature comparable with *3d* orbital which is origin of typical magnetic transition metals.[34] Thus, if the D($E_f$) is large enough, Stoner criterion can be satisfied (see Figure S3). On the other hand, similar to n-type interface, the stability of ferromagnetic ordering also depends on the polarization strength of bulk $PbTiO_3$. As shown in Figure 7(a), there is no energy difference between FM and AFM ordering when the polarization strength is small and the interface is paramagnetic without persistent spin-polarization. When the ferroelectric polarization increases above ~0.6 C/m$^2$, the conductive state becomes spin-polarized and the FM ordering is energetically more favorable than the AFM ordering. The magnitude of the magnetic moment and the charge carrier density of this p-type conductive state at the interface are also sensitive to the ferroelectric polarization and can be tuned in a wide range (Figure 7 (a) and (b)). As a results, the carrier type, density and magnetic properties at the interface are all determined by the polarization of PbTiO$_3$, which can be controlled by electric field or strain.

In conclusion, we propose a new type of oxide heterostructure consisting two insulating nonpolar and nonmagnetic oxides that hosts a spin-polarized ferromagnetic state at the interface due to the polarization discontinuity. Using first principles calculations, we investigate the concrete example of TiO$_2$/PbTiO$_3$ heterostructure. The calculation results show that the FM ordering is stable for this interfacial magnetism due to partially occupied Ti-3$d$ orbitals at n-type interface or O-2$p$ orbitals at p-type interface. The spontaneous magnetic ordering can be understood by the Stoner picture. Importantly, the key properties of the interfacial state including the carrier type, density and magnetic moment are sensitive to the ferroelectric polarization of PbTiO$_3$ hence can be conveniently controlled by applied electric field or by strain. Our discovery provides an alternative way to generate 2D carriers other than polar discontinuity. The TiO$_2$/PbTiO$_3$ prototype system shows several advantageous aspects over the LaAlO$_3$/SrTiO$_3$ interface, including switchable carrier types, high carrier density, robust ferromagnetism, and most importantly, the controllability through the ferroelectric polarization. The coupling between the ferroelectric polarization and the interfacial ferromagnetism points to an intriguingly novel interfacial multiferroic system. The results are expected to open a new avenue to design 2D carriers at oxide interface with large magnetoelectric effects, which is promising for spintronic applications and the emergence of novel quantum phases.

**Experimental Section**

Our first-principles calculations are based on the density functional theory (DFT) within generalized gradient approximation (GGA) with PBEsol[35] formula implemented in the Vienna Ab-initio Simulation Package.[36] The projector augmented wave pseudopotential method is employed to model ionic potentials.[37] Kinetic energy cutoff is set above 400 eV for all calculations. Monkhorst-Pack k-point sampling[38] is used for the Brillouin zone integration: $10 \times 10 \times 10$ for bulk $PbTiO_3$, $10 \times 10 \times 4$ for bulk $TiO_2$, and $10 \times 10 \times 1$ for the $TiO_2/PbTiO_3$ heterostructure. Bulk $PbTiO_3$ and $TiO_2$ are fully optimized with respect to the ionic positions until the forces on all atoms are less than 0.01 eV/Å. To study the electronic properties of the interface of $TiO_2/PbTiO_3$ heterostructure, a symmetric $PbTiO_3/TiO_2/PbTiO_3$ slab model is used in order to maintain periodic boundary condition without the dipole correction. The slab contains 13 ($TiO_2$) layers sandwiched by five $PbTiO_3$ unit cell layers on each side. The $TiO_2$ and $PbTiO_3$ thickness are tested to ensure that the distance between two interfaces and surfaces is large enough such that the exchange interactions as well as the excitonic binding between them can be disregarded. The ferroelectricity of $PbTiO_3$ is considered by fixing the atomic positions of the $PbTiO_3$ slab to their bulk positions except for the unit neighboring the interface. This is reasonable as the thickness of the $PbTiO_3$ substrate is normally larger than 100 nm in experiments, so the bulk properties are dominant. In addition, it has been reported that a transition from a polydomain to monodomain ferroelectric tetragonal phase occurs at room temperature when the $PbTiO_3$ thickness $d \geq 2$ nm (~five unit cells).[39] Because of the polarization discontinuity, some superficial free charges are required to accumulate at the $PbTiO_3$ free surface facing the vacuum, which are sources of screening that can help to stabilize monodomain phases in ferroelectric thin films and are mostly provided by surface redox processes (like formation of charged defects or adsorption of chemical species) in experiments. In our calculation, the minimum vacuum layer thickness is greater than 20 Å, which is large enough to avoid artificial interaction between neighboring images. To appropriately describe the electronic states of the correlated Ti *3d* electrons, the on-site Hubbard U parameter is tested and it is found that the magnetic properties are quite stable with varying U parameter (Figure.S4). Thus, a moderate value of effective U=3 eV for the Ti *3d* states is used for the results to be presented in the following.

**ACKNOWLEDGMENTS**

We thank Z.L. Zhang for valuable discussions. This work was supported by the National Natural Science Foundation of China (Grant No. 61574123, 11374009, 11474249, 51232006 and 51472218), National Key R&D Program of the MOST of China (Grant No. 2016YFA0300204), the National 973 Program of China (2015CB654901), National Young 1000 Talents Program of China, SUTD-ZJU Collaborative Research Grant (SUTD-ZJU/PILOT/01/2015), Fundamental Research Funds for the Central Universities (2016FZA4005) and Special Program for Applied Research on Super Computation of the NSFC-Guangdong Joint Fund (the second phase). K. Yang acknowledges the start-up funds from the University of California, San Diego.

# Figures

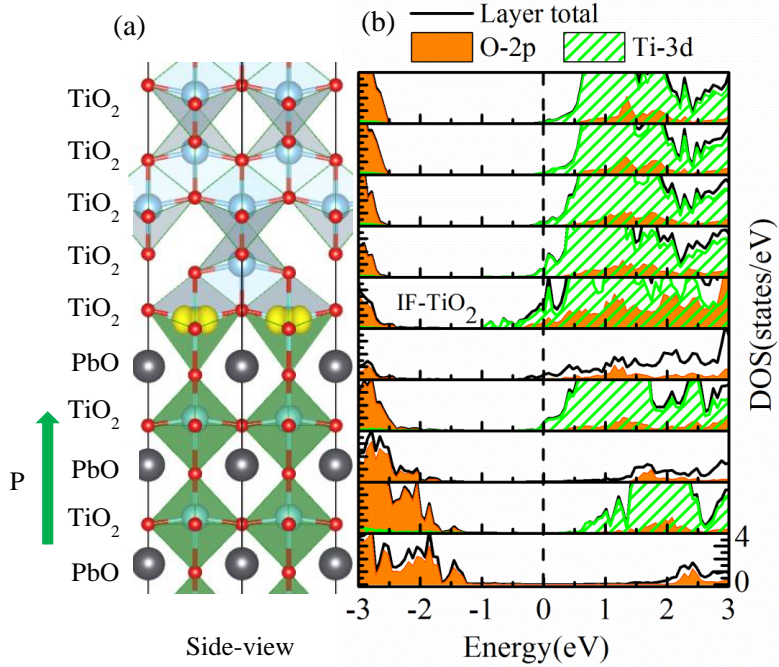

Figure 1. (Color online) The interface structure of $TiO_2/PbTiO_3$ along with the spin density near the interface region (a), calculated magnetic ground state layer-resolved partial DOS for $TiO_2/PbTiO_3$ (b). The red, light blue and black solid balls represent the O, Ti and Pb atoms respectively. The isosurface of spin density is plotted with yellow shadow. The green arrow indicates the spontaneous polarization direction of $PbTiO_3$. The vertical line in (b) denotes the Fermi energy ($E_f$). These notations represent the same means respectively in the follow part.

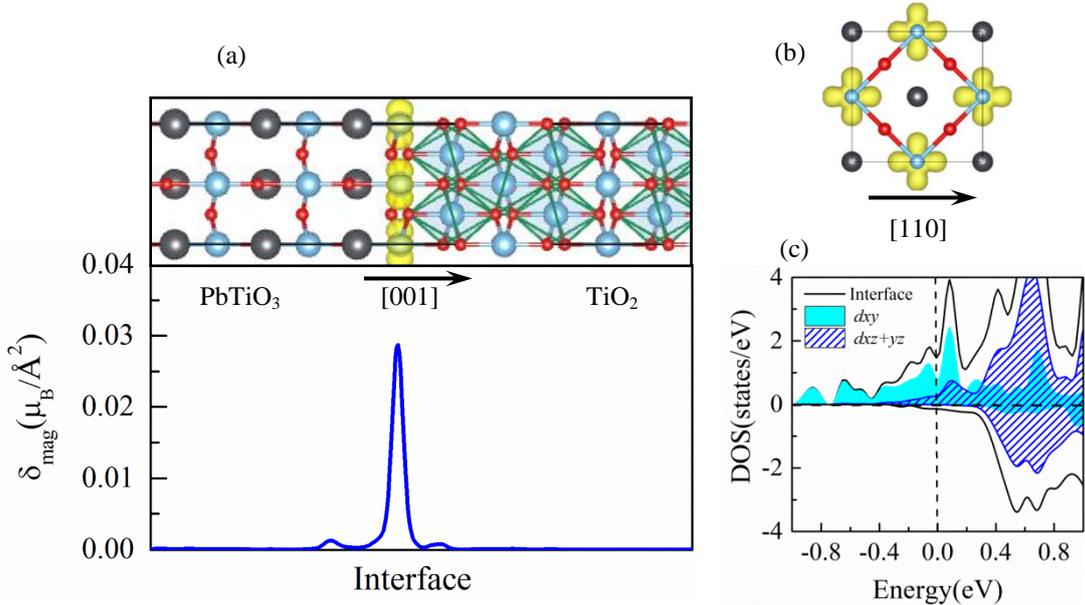

Figure 2. (Color online) (a) Plane-averaged magnetization density for the HS at polarization P= $0.72 C/m^2$. (b) The spin density character in the interface $TiO_2$ layer. (c) DOS for interface $TiO_2$ layer, as well as resolved $d_{xy}$ and $d_{xz+yz}$ components for Ti-3d. The vertical line in (c) denotes the Fermi energy.

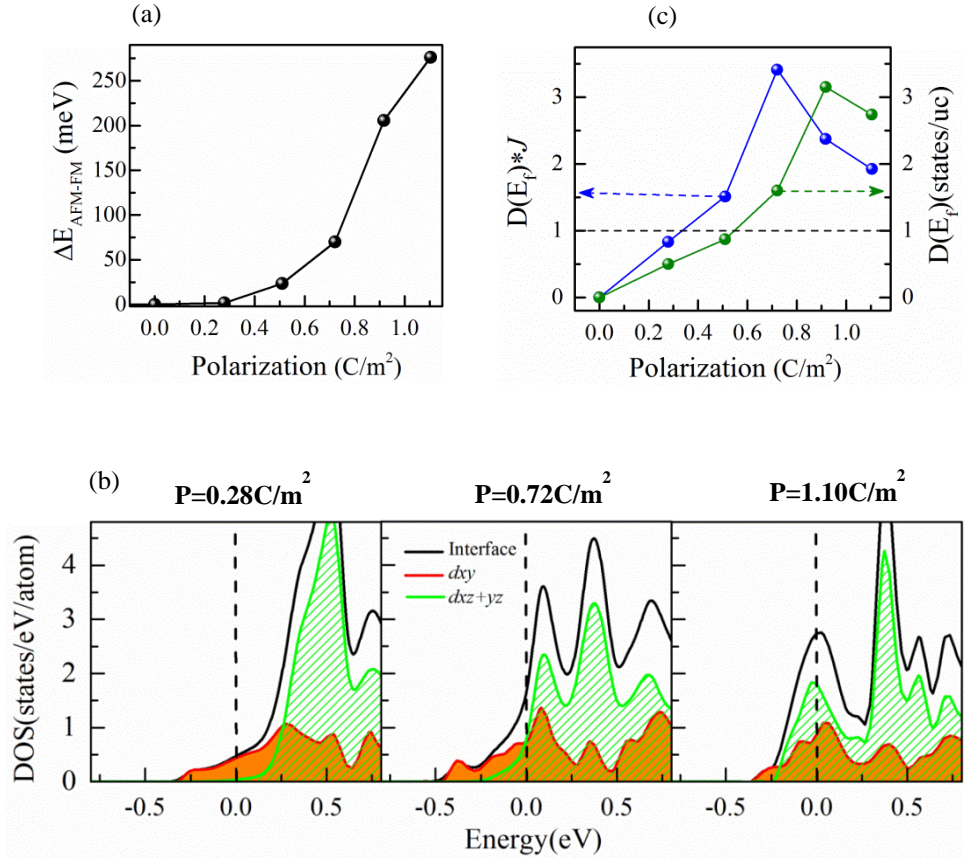

Figure 3. (Color online) (a) Energies of C-AFM state relative to the FM state with different polarization. (b) Nonmagnetic PDOS resolved into $d_{xy}$ and $d_{xz+yz}$ components projected onto the interface $TiO_2$ layer, for systems with polarizations P=0.28, 0.72, and 1.10 C/m$^2$ from left to right. The vertical line denotes the Fermi energy. (c) Stoner parameter and the density of states around Fermi lever as a function of the polarization.

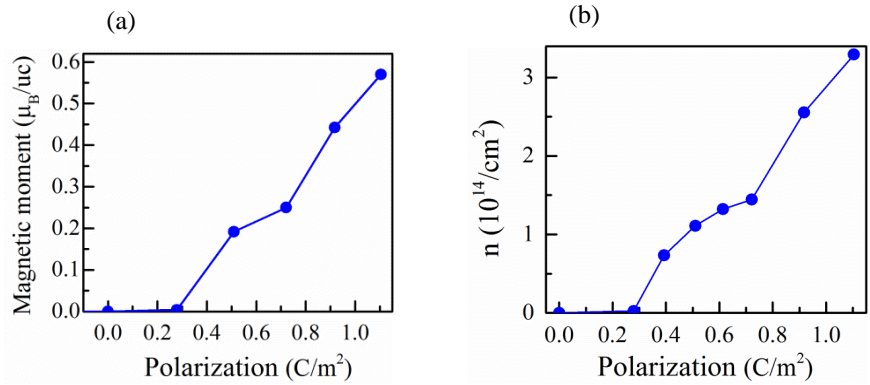

Figure 4. The magnetic moment per unit cell (uc) (a), n-type interfacial charge carrier density (b), with respect to the polarization.

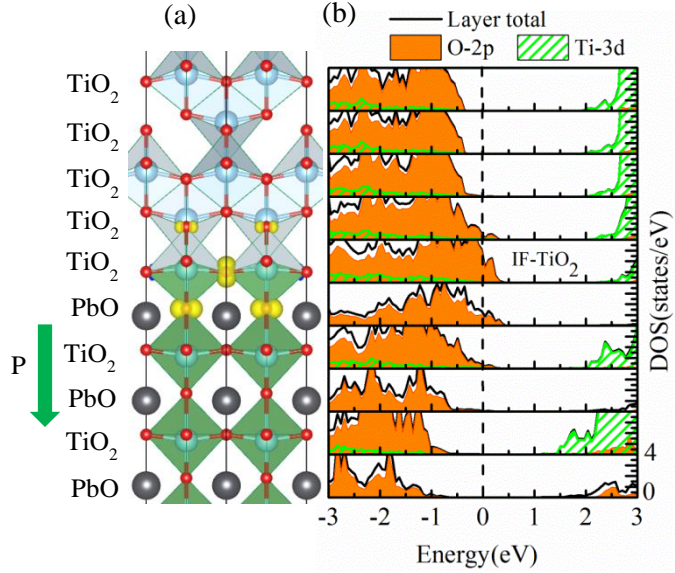

Figure 5. (Color online) The interface structure of $TiO_2/PbTiO_3$ along with the spin density near the interface region (a), calculated magnetic ground state layer-resolved partial DOS for $TiO_2/PbTiO_3$ (b). The red, light blue and black solid balls represent the O, Ti and Pb atoms respectively. The yellow shadow represents the isosurface of spin density. The green arrow indicates the spontaneous polarization direction of $PbTiO_3$.

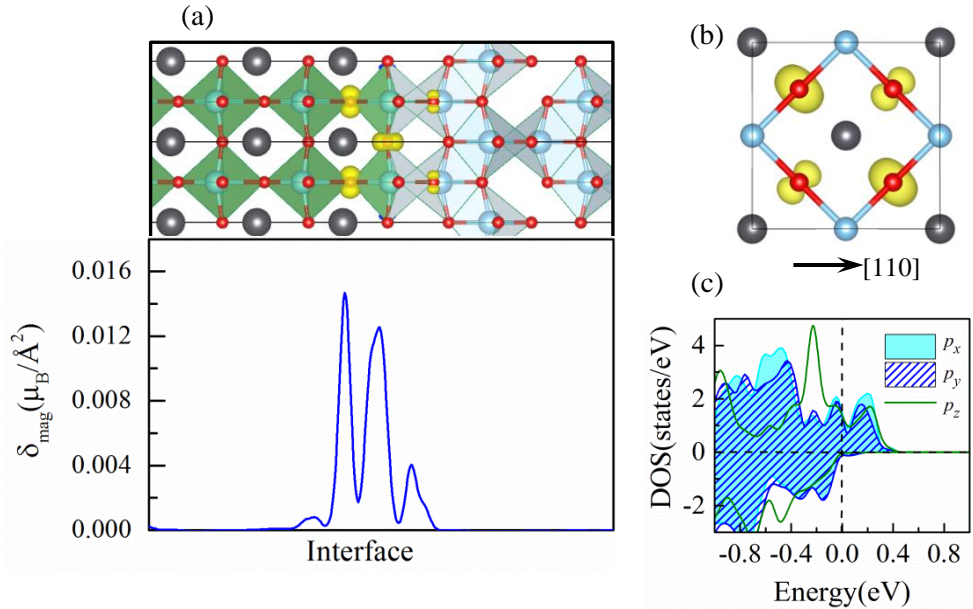

Figure 6. (Color online) Plane-averaged magnetization density for the HS at normal polarization (P=0.72C/m$^2$) (a), the top view of the spin density near the interface region (b), PDOS for the interface layer, O-2$p$ orbital is resolved into $p_x$, $p_y$ and $p_z$ (c).

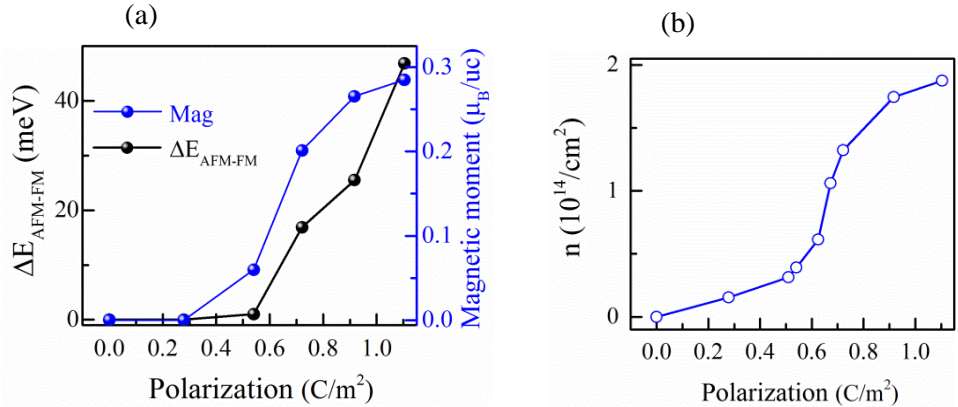

Figure 7. (Color online) Energies of AFM state relative to the FM state (the left axis), and the magnetic moment per unit cell (uc) (the right axis) (a), p-type interfacial charge carrier density (b), with respect to the polarization.